\documentstyle[12pt]{article}
\begin{document}
\title{{\bf No-Bang Quantum State of the Cosmos}
\thanks{Alberta-Thy-08-07, arXiv:0707.2081}}
\author{
Don N. Page
\thanks{Internet address:
don@phys.ualberta.ca}
\\
Institute for Theoretical Physics\\
Department of Physics, University of Alberta\\
Room 238 CEB, 11322 -- 89 Avenue\\
Edmonton, Alberta, Canada T6G 2G7
}
\date{(2008 March 7)}

\maketitle
\large
\begin{abstract}
\baselineskip 18 pt

	A quantum state of the entire cosmos (universe or multiverse)
is proposed which is the equal mixture of the Giddings-Marolf states
that are asymptotically single de Sitter spacetimes in both past and
future and are regular on the throat or neck of minimal
three-volume.  That is, states are excluded that have a big bang or
big crunch or which split into multiple asymptotic de Sitter
spacetimes.  (For simplicity, transitions between different values of
the cosmological constant are assumed not to occur, though different
positive values are allowed.)  The entropy of this mixed state
appears to be of the order of the three-fourths power of the
Bekenstein-Hawking $A/4$ entropy of de Sitter spacetime.  Most of the
component pure states do not have rapid inflation, but when an
inflaton is present and the states are weighted by the volume at the
end of inflation, a much smaller number of states may dominate and
give a large amount of inflation and hence may agree with
observations.

\end{abstract}
\normalsize

\baselineskip 13.9 pt

\newpage

\section*{Introduction}

	Our observations strongly suggest that our observed portion
(or subuniverse \cite{Wein87} or bubble universe \cite{Linde96,Vil97}
or pocket universe \cite{Guth00}) of the entire universe (or
multiverse
\cite{James,Lodge,Leslie,Gell-Mann,Deutsch,Dyson,Rees,Carr} or
metauniverse \cite{Vil95} or omnium \cite{Penrose-book} or megaverse
\cite{Susskind-book} or holocosm \cite{Page-in-Carr}, though I am
henceforth abandoning this neologism after my wife warned me recently
that it might potentially be interpreted as offensive because of its
similarity to holocaust, which was certainly not my intent, and I
apologize to anyone whom I might have offended) is much more special
than is implied purely by the known dynamical laws.  For example, it
is seen to be enormously larger than the Planck scale, with small
large-scale curvature, and with approximate homogeneity and isotropy
of the matter distribution on the largest scales that we can see
today.  It especially seems to have had extraordinarily high order in
the early universe to enable its coarse-grained entropy to increase
and to give us the observed second law of thermodynamics
\cite{Tolman,Davies,Penrose-in-HI}.

	Leading proposals for special quantum states of the universe
have been the Hartle-Hawking `no-boundary' proposal
\cite{HVat,HH,H,HalHaw,P,Hal,HLL,HH1,Page-Hawking-BD,HH2,HHH} and the
`tunneling' proposals of Vilenkin, Linde, and others
\cite{V,TL,TZS,TR,TVV,TGV}. In simplified toy models with a suitable
inflaton, both of these classes of models have seemed to lead to the
special observed features of our universe noted above.

	However, Leonard Susskind \cite{Susspriv} (cf.\
\cite{DKS,GKS,Sus03}) has made the argument, which I have elaborated
\cite{DP2006}, that in the no-boundary proposal the cosmological
constant or quintessence or dark energy that is the source of the
present observations of the cosmic acceleration
\cite{Perl,Riess,PTW,Tonry,WMAP,Tegmark,Astier} would give a large
Euclidean 4-hemisphere as an extremum of the Hartle-Hawking path
integral that would apparently swamp the extremum from rapid early
inflation.  Therefore, to very high probability, the present universe
should be very nearly empty de Sitter spacetime, which is certainly
not what we observe.

	The tunneling proposals have also been criticized for various
problems \cite{BH,GV97,Lin98,HT,TH,V98}.  For example, the main
difference from the Hartle-Hawking no-boundary proposal seems to be
the sign of the Euclidean action \cite{TL,V}.  It then seems
problematic to take the opposite sign for inhomogeneous and/or
anisotropic perturbations without leading to some instabilities, and
it is not clear how to give a sharp distinction between the modes
that are supposed to have the reversed sign of the action and the
modes that are supposed to retain the usual sign of the action. 
Vilenkin has emphasized \cite{V,TGV} that the instabilities do not
seem to apply to his particular tunneling proposal, which does not
just reverse the sign of the Euclidean action.  However, Vilenkin
admits \cite{TGV} that ``both wavefunctions are far from being
rigorous mathematical objects with clearly specified calculational
procedures.  Except in the simplest models, the actual calculations
of $\psi_{\mathrm T}$ and $\psi_{\mathrm HH}$ involve additional
assumptions which appear reasonable, but are not really well
justified.''

	Therefore, at least unless and until any of these proposals
can be made rigorous and can be shown conclusively to avoid the
problems attributed to them, it is worth searching for and examining
other possibilities for the quantum state of the universe or
multiverse.  In this paper a schematic proposal for another state is
made for universes with positive cosmological constant, based upon
the global picture of `eternal de Sitter space' given recently by
Giddings and Marolf \cite{GM}.  They assert, ``Eternal de Sitter
physics is described by a finite dimensional Hilbert space in which
{\it each} state is precisely invariant under the full de Sitter
group.''  Here I make a further restriction of the states and then
propose that the quantum state of the universe is the uniform mixture
of all of the resulting finite set of pure states.  Since none of
these pure states have the entire universe beginning at a big bang or
ending at a big crunch, I call the resulting mixed state the `no-bang
state.'  This state appears to be a quantitative improvement over the
no-boundary proposal, but unfortunately it still seems to suffer from
a similar qualitative problem.

	For simplicity, I shall assume that asymptotically de Sitter
spacetimes are absolutely stable, with no transitions between
different values of the cosmological constant, though if different
positive values of the cosmological constant are allowed, the no-bang
state includes all of them.  I shall neglect zero and negative values
of the cosmological constant and therefore exclude states in which
they occur (though I am not opposed to a possible extension of the
no-bang state that includes components with zero and negative
cosmological constant, if one can make some definite proposal for
them).

\section{Giddings-Marolf eternal de Sitter states}

	Giddings and Marolf \cite{GM} show how one may use group
averaging to convert a dS-noninvariant `seed' state of matter to a
dS-invariant state, as all physical states must be in quantum
gravity.  (Actually, all physical states must be invariant under the
full diffeomorphism group, but the de Sitter group is a crucial
subgroup of this when there is a positive cosmological constant, as
is assumed here.)  Giddings and Marolf then consider seed states with
weak back-reaction and focus on quasiclassical seed states in each of
which the stress-energy tensor has fairly definite values at each
point in the background de Sitter space.  (After the group averaging,
the resulting dS-invariant physical state will give an identical
superposition of stress-energy tensor values at all points of the
spacetime and in all local Lorentz frames, so it is only the seed
states that can be inhomogeneous quasiclassical states in the sense
of having fairly definite field values that vary over the spacetime.)

	Giddings and Marolf \cite{GM} choose a particular foliation
of de Sitter space (say 4-dimensional, as I shall assume for
simplicity here, though the arguments may be modified simply for any
higher dimension) by hypersurfaces that are 3-spheres of symmetry. 
The sequence of these hypersurfaces shrinks down from asymptotically
infinite size at past timelike infinity to a `neck' of minimal size
and then re-expands back to asymptotically infinite size at future
timelike infinity.  Since the background de Sitter space is globally
hyperbolic, the seed state over the entire space will be causally
determined by its restriction to the neck (at least when one ignores
gravity, and presumably even when gravity is included, as shall be
assumed here).  This restriction to the minimal 3-sphere neck can be
considered to be the `initial' data for the seed state, though of
course it is temporally before just the part of the spacetime to the
future of the neck.  (For the part of the spacetime to the past of
the neck, this `initial' state might instead be considered a `final'
state, though it is not the final hypersurface on which the seed
state may be evaluated either.)

	Giddings and Marolf \cite{GM} consider seed states which have
small gravitational back-reaction at the neck.  By calculating the
entropy of thermal massless radiation of energy density less than
that of the cosmological constant $\Lambda$, they estimate that the
logarithm of the number of such states is, in 4 dimensions, roughly
$(R/\ell_p)^{3/2}$ (a quantity which had also been obtained earlier
by Banks, Fischler, and Paban \cite{BFP}), where $R$ is the radius of
the neck (given by $b \equiv \sqrt{3/\Lambda}$ when there is
negligible gravitational back-reaction), and $\ell_p$ is the Planck
length.  This logarithm is much less than the Bekenstein-Hawking
entropy of de Sitter space, which is $\pi(R/\ell_p)^2$, one quarter
the area of the cosmological horizon in Planck units ($\hbar = c = G
= k_{\mathrm Boltzmann} = 1$).  The much larger latter entropy
suggests \cite{GM} that there are far more non-perturbative
gravitational states, perhaps black holes of very large size,
comparable to $R$.

	Since states with black holes of very large size do not seem
to fit our observations of the universe, I would like to exclude them
from the mixed state I am proposing.  However, I would prefer not
simply to restrict to states of small gravitational back-reaction,
first because it is rather arbitrary how to define ``small,'' and
second because I would not want to exclude stellar-mass black holes in
the present universe, which certainly do not have small gravitational
back-reaction.

\section{No-bang seed states}

	My proposal is to consider seed states for which one has a
regular neck, and for which the seed state evolves to a single
asymptotically de Sitter space in both the infinite past and the
infinite future, without the entire universe having either a big bang
in the past or a big crunch in the future, and without the universe
splitting up into more than one asymptotically de Sitter space in
either the infinite past and the infinite future as it would do for
necks with large black holes.  However, no restriction is proposed
that gravity need be weak on the neck, or that no black holes occur
either to the past or the future of this neck.  (I shall assume that
all holes do decay away in both the asymptotic past and future, say by
Hawking radiation in the future and its time reverse in the past.)

	Here the neck is defined as a complete closed 3-dimensional
locally-minimal hypersurface (a complete closed hypersurface with
zero mean extrinsic and with smaller 3-volume than any other nearby
complete closed hypersurface with constant mean extrinsic curvature),
the hypersurface of globally minimal 3-volume if more than one such
locally-minimal hypersurfaces occur.  Since my discussion will be
almost entirely at the semiclassical level, I shall assume that each
seed state gives a spacetime geometry that is sufficiently
quasiclassical, at least in the neighborhood of the neck, that one
can get a good approximate description of the spacetime geometry in
that region.  (Elsewhere in the spacetime, such as where quantum
uncertainties in gravitational collapse may occur, as in the
formation and evaporation of black holes, one need not be able to
describe the seed state as giving a single approximately classical
geometry.)

\section{Homogeneous, isotropic necks}

	First I shall consider the case in which the neck is
approximately a homogeneous, isotropic 3-sphere of radius $R$, which
Giddings and Marolf \cite{GM} used to estimate the entropy of seed
states of weak gravitational back-reaction.  For now I shall follow
them in assuming that the perturbations from homogeneity and
isotropy are small, but I shall allow arbitrarily large (though
nearly homogeneous and isotropic) matter stress tensors and
arbitrarily large back-reaction upon the size of the neck and the
subsequent evolution of the spacetime.

	I shall use quantum units in which $\hbar = c = k_{\mathrm
Boltzmann} = 1$, but since there are different conventions as to
whether the Planck length $\ell_p$ is $G^{1/2}$ (my usual
convention) or $(8\pi G)^{1/2}$, I shall in this paper explicitly
include the factors of Newton's gravitational constant $G$.

	I shall assume that there is a positive cosmological
constant $\Lambda = 3/b^2 \ll G^{-1}$ with characteristic length
scale much larger than the Planck length, $b \gg G^{1/2}$ (the
radius of the resulting de Sitter spacetime if no matter or
gravitational waves were present).  I shall assume that the matter
most relevant near the neck consists of an inflaton field plus
thermal radiation.  For simplicity, I shall assume that the inflaton
scalar field $\phi$ is a homogeneous free massive scalar field of
mass $m$ whose Compton wavelength is much less than $b$ but much
greater than the Planck length, $\Lambda \ll m^2 \ll G^{-1}$.  Its
energy density is $\rho_\phi = \frac{1}{2}\dot{\phi}^2 + \frac{1}{2}
m^2 \phi^2$, and its isotropic pressure is $P_\phi =
\frac{1}{2}\dot{\phi}^2 - \frac{1}{2} m^2 \phi^2$, where an overdot
denotes a derivative with respect to the proper time that is
measured normally to the homogeneous, isotropic 3-spheres of radius
$R(t)$ that foliate the universe near the neck ($t=0$).  I shall
assume that the thermal radiation is also homogeneous and isotropic
and is dominated by fields whose mass can be neglected, so that its
energy density is $\rho_{\mathrm rad} = aT^4$ and its isotropic
pressure is $P_{\mathrm rad} = (1/3)aT^4$, where $T$ is the
temperature (which evolves proportional to $1/R$ away from the
neck), and $a$ is the radiation constant (e.g., $a = 427\pi^2/120$
for the fields of the Standard Model of particle physics at high
temperature \cite{PDG}).

	Thus I assume that near the neck, the spacetime geometry is
given by a closed ($k=+1$) FRW model with time-dependent radius
$R(t)$.  The Einstein constraint equation is
\begin{equation}
H^2 \equiv \left(\frac{\dot{R}}{R}\right)^2
=\frac{4\pi G}{3}(m^2\phi^2+\dot{\phi}^2+2aT^4)
+\frac{\Lambda}{3}-\frac{1}{R^2},
\label{eq:1}
\end{equation}
and the equation of evolution is
\begin{equation}
\frac{\ddot{R}}{R}
=\frac{4\pi G}{3}(m^2\phi^2-2\dot{\phi}^2-2aT^4)
+\frac{\Lambda}{3}.
\label{eq:2}
\end{equation}

	The neck has $\dot{R}=0$, so at the neck
\begin{equation}
R = \left(\frac{4\pi G}{3}\right)^{-1/2}
\left(m^2\phi^2+\dot{\phi}^2+2aT^4+\frac{\Lambda}{4\pi G}
\right)^{-1/2}.
\label{eq:3}
\end{equation}
For the neck to be a minimal 3-sphere, one needs $\ddot{R}\geq 0$,
which puts a limitation on $\dot{\phi}$ and $T$.  This has the
implication that the minisuperspace constrained phase space
\cite{GHS} for the FRW model with the cosmological constant and the
inflaton without the thermal radiation is $2\pi^2 R^3 d\phi\wedge
d\dot{\phi}$, which when using Eq. (\ref{eq:1}) integrates to a
finite total measure over this phase space, under the condition that
there be a neck.

	When one puts the radiation back in, most of the states will
come from the entropy of the radiation.  For fixed $\phi$, the
maximum entropy of the radiation will come from having the maximum
value of the temperature $T$ consistent with $\ddot{R}\geq 0$ at
$\dot{\phi}=0$, which gives
\begin{equation}
T = (2a)^{-1/4}\left(m^2\phi^2+\frac{\Lambda}{4\pi G}\right)^{1/4},
\label{eq:4}
\end{equation}
\begin{equation}
R = \left(\frac{8\pi G}{3}\right)^{-1/2}
\left(m^2\phi^2+\frac{\Lambda}{4\pi G}\right)^{-1/2}.
\label{eq:5}
\end{equation}

	Since the 3-volume of the neck is $V = 2\pi^2 R^3$, and since
the entropy of the radiation is $S = (4/3)aT^3V$, one finds that the
maximum entropy for each value of the inflaton field $\phi$ at the
neck is
\begin{equation}
S = \left(\frac{9\pi^2 a}{512}\right)^{1/4}
\left(G^2m^2\phi^2+\frac{G\Lambda}{4\pi}\right)^{-3/4}.
\label{eq:6}
\end{equation}
For example, for $\phi=0$, $T = [\Lambda/(8\pi G a)]^{1/4} = [3/(8\pi
G a)]^{1/4}b^{-1/2}$ and $R = (2\Lambda/3)^{-1/2} = b/\sqrt{2}$
(slightly smaller than the de Sitter radius $b = \sqrt{3/\Lambda}$ at
$T=0$, since the maximal thermal radiation for $\ddot{R}\geq 0$
shrinks the size of the neck), so
\begin{equation}
S = S_{\mathrm no-bang}
= \left(\frac{9\pi^5 a}{8G^3\Lambda^3}\right)^\frac{1}{4}
= [\pi^5 a/(24 G^3)]^{1/4}b^{3/2}
= [\pi^5 a/(3 G^3)]^{1/4}R^{3/2},
\label{eq:7}
\end{equation}
agreeing with the order-of-magnitude estimate of Giddings and Marolf
\cite{GM} for the maximal entropy of states with weak gravitational
back-reaction.  Here we get the same order-of-magnitude limitation as
they did, but now for the entropy of all states with approximate
3-sphere symmetry at the neck, whether or not the homogeneous,
isotropic stress-energy tensor of the inflaton plus radiation is a
weak perturbation of de Sitter spacetime.  (However, see below
for how highly anisotropic necks can have arbitrarily large radiation
entropy, though they will not evolve to single asymptotically de
Sitter regions.)

	If we take $\Omega_\Lambda = 0.72\pm 0.04$ from the
third-year WMAP results of \cite{WMAP} and $H_0 = 72\pm 8$ km/s/Mpc
from the Hubble Space Telescope key project \cite{Freedman}, and drop
the error uncertainties, we get $G\Lambda = 3\Omega_\lambda H_0^2
\approx 3.4\times 10^{-122}$, which would give $b = \sqrt{3/\Lambda}
\approx 9.4\times 10^{60}\sqrt{G}$.  This then gives an entropy of
the part of the no-bang mixed state corresponding to our value of the
cosmological constant of $S_{\mathrm no-bang} \approx 4\times
10^{90}$.

	Although this is very roughly the same value as the entropy
in the cosmic microwave background photons within the observable
universe (by the perhaps anthropically explained coincidence that the
photon energy density today is within a few orders of magnitude of
the cosmological constant), conceptually this is not the same
quantity. On one hand, it represents the von Neumann entropy of the
entire spacetime that might be much larger than that of our observed
region.  On the other hand, the entropy in the cosmic microwave
background photons within the observable universe is a coarse-grained
entropy and also might be much larger than the fine-grained von
Neumann entropy of the same region.

	For large values of the inflaton at the neck, $m^2\phi^2 \gg
\Lambda/(4\pi G)$, although the maximum temperature rises as
$\phi^{1/2}$, the 3-volume of the neck shrinks as $\phi^{-3}$, so the
radiation entropy, $S = (8\pi^2 a/3)(RT)^3$, shrinks as
$\phi^{-3/2}$.  Therefore, most of these seed states with a nearly
homogeneous, isotropic neck have only low values of the inflaton
field and hence negligible amounts of (rapid) inflation from the
inflaton (not counting the much slower asymptotic exponential
expansion from the cosmological constant $\Lambda$ itself as
inflation).  That is, for these `eternal de Sitter space' states,
inflation is highly improbable, analogous to the restricted class of
minisuperspace states used in \cite{GT} but different from the
ambiguous probability of inflation for all minisuperspace states
\cite{HP88}, where both the inflationary and non-inflationary
solutions have infinite measure.

	On the other hand, if one looks at this FRW universe not at
the neck but at a time after the possible rapid inflation (say at the
end of inflation, or at some fixed value of the matter energy density
that is lower than $m^2$ to avoid looking during inflation), then the
volume of space will be much larger if rapid inflation has occurred. 
Therefore, if one wants the expectation value of the volume of space
at the end of inflation, one should weight each state by the volume
then.  A motivation for doing this would be Vilenkin's Principle of
Mediocrity \cite{Vil95} that we are a typical civilization, along
with the assumption that the number of civilizations would be
proportional to the volume of space at the end of inflation.  Other
motivations are given in \cite{Veternal,AV,Leternal,Lindebook,LLM,
Seternal,Geternal,GSVW,Hawvol,HHH}, though there is also
disagreement \cite{GT}.

	If we start with a value of the free massive inflaton field
that gives $\phi^2 \gg G^{-1} \gg \Lambda/(G m^2)$, then most of the
rapid inflation will occur during the slow roll phase in which $H
\approx \sqrt{4\pi G/3}\;m\phi$, and the Klein-Gordon equation for
the inflaton, $\ddot{\phi} + 3H\dot{\phi} + m^2\phi = 0$, will give
$\dot{\phi} \approx -m^2\phi/(3H)$ and hence $dt \approx -3H
d\phi/(m^2\phi)$, so the number of e-folds of inflation from some
large initial $\phi$ to the end of inflation at $\phi \sim G^{-1/2}$
will be \cite{Lindebook,LL} \begin{equation} N = \int Hdt \approx
2\pi G \phi^2. \label{eq:8} \end{equation} Therefore, the volume of
space at the end of inflation with a free massive inflaton of initial
value $\phi$ will be $V \sim (m\phi)^{-3}e^{3N} \sim \exp{(6\pi G
\phi^2)}$, dropping the prefactor in the final expression as having a
logarithm much closer to zero than the enormous exponential.  Here,
to get finite answers, I initially ignore quantum fluctuations of the
inflaton field during inflation that might lead to an arbitrarily
large volume from eternal inflation
\cite{Veternal,AV,Leternal,Lindebook,LLM,Seternal,Geternal,GSVW}.

	When one uses this slow-roll inflationary volume $V \sim
\exp{(6\pi G \phi^2)}$ along with $e^S$ for the number of states, the
resulting expectation value of the volume at the end of inflation
would be roughly
\begin{equation}
\langle V\rangle \sim \frac{\int Ve^S d\phi}{\int e^S d\phi}.
\label{eq:9}
\end{equation}

It would be somewhat better to include the minisuperspace constrained
phase-space factor \cite{GHS} $2\pi^2 R^3 d\phi\wedge d\dot{\phi}$,
which by itself integrates to a finite value when one includes the
restriction $\ddot{R}\geq 0$ that limits $\dot{\phi}$ at the neck, as
discussed above, making the denominator of the right hand side of Eq.
(\ref{eq:9}) finite, even after integrating over an infinite range of
$\phi$.  However, since $V$ grows so rapidly with $\phi$ (the value
of the instanton field at the neck), qualitatively it is not very
important to be so refined.

	If one takes the integrals in Eq. (\ref{eq:9}) to be over an
infinite range of $\phi$, then the numerator will diverge violently,
so that $\langle V\rangle$ will be infinite.  To get a finite answer,
one might cut off the integral at a value of $\phi$ corresponding to,
say, $\epsilon$ times the Planck energy density $G^{-2}$, where
$\epsilon$ is some number of order unity, so $(1/2)m^2\phi^2 <
\epsilon G^{-2}$.  Then the slow-roll inflationary volume will be $V
< \exp{(12\pi \epsilon G^{-1} m^{-2})}$.  If one uses the estimate
that $m \approx 1.5\times 10^{-6} G^{-1/2} \approx 7.5\times 10^{-6}
(8\pi G)^{-1/2}$ \cite{Lindebook,LL} from the measured fluctuations
of the cosmic microwave background, one then gets $V <
\exp{(1.7\times 10^{13}\epsilon)}$.

	Since the entropy $S$ of the thermal radiation at the large
value of $\phi$ at the neck needed to give this maximal slow-roll
inflationary value is of the order of unity, the contribution to the
integral in the numerator of Eq. (\ref{eq:9}) near the maximum of the
restricted range of $\phi$ is of very roughly of the order of
$\exp{(1.7\times 10^{13}\epsilon)}$.  On the other hand, the
contribution to the integral from small values of the inflaton field
$\phi$, say $m^2\phi^2 < \Lambda/G$, will have huge radiation entropy
values, $S \approx S_{\mathrm no-bang} \approx 4\times 10^{90}$, so
this region of the integral will give a contribution that is of very
roughly of the order of $\exp{(4\times 10^{90})}$, which is far, far
larger than the contribution near the maximum value of $\phi$ (if the
maximum inflaton field is indeed limited by energy densities near the
Planck value, i.e., for $\epsilon \ll 10^{77}$).

	Therefore, even if we weight by the 3-volume $V$, say at some
fiducial energy matter energy density less than $\Lambda/8\pi G$ in
order to have this energy density possible even without inflation,
the non-inflationary solutions overwhelmingly dominate, when we use
the slow-roll inflation value $3N \approx 6\pi G \phi^2$ for the
logarithm of the growth factor of the volume of the universe during
inflation from a large initial value $\phi$.

	To avoid this conclusion for the no-bang quantum state, we
can appeal to eternal inflation
\cite{Veternal,AV,Leternal,Lindebook,LLM,Seternal,Geternal,GSVW},
which predicts that quantum fluctuations of the inflaton field will
lead to an arbitrarily large amount of inflation and hence $\langle
V\rangle = \infty$.  Then one would indeed get that the inflationary
solutions dominate when one weights by the volume.

	In order that eternal inflation occur with high probability
\cite{Veternal,AV,Leternal,Lindebook,LLM,Seternal,Geternal,GSVW}, one
needs the inflaton to start in the quantum diffusion regime, which
for a free massive scalar inflaton gives $\phi > \phi_{\mathrm q}
\sim G^{-3/4} m^{-1/2} \sim 10^3 G^{-1/2}$.  Restricting $\phi$ in
this way on the neck then restricts the entropy of the thermal
radiation on the neck to be $S \sim (Gm^2)^{-3/8} \sim 10^{9/2}$. 
Thus we would need a few tens of thousands of bits of information to
say which of the pure states in the no-bang mixed state with $\phi >
\phi_{\mathrm q}$ we are in.  However, because eternal inflation
would spread this information over an enormous volume, we would be
unlikely ever to be be able to determine it.

	In comparison, the logarithm of the number of possible
sequences of the roughly 3 billion DNA base pairs in the human genome
is $\ln{(4^{3\times 10^9})} \sim 4\times 10^9$, around a hundred
thousand times greater than the eternal inflation universe entropy
calculated above.  In this sense, the information needed to specify
which of these $e^S$ pure states one is in is hundreds of thousands
of times less than the information needed to specify one's genome.

	(This estimate does not consider possible compression of the
information.  Since one would expect that only a very tiny fraction
of genomes of 3 billion base pairs would be viable, the compressed
information to specify which viable genome one has is likely to be
many times smaller than the 6 billion bits just needed to list the
sequence.  It is surely an open question whether the human genome
information could be compressed down to several tens of thousands of
bits, comparable to the uncompressed information to specify which of
the $e^S$ pure states we are in, which might also be compressible.)

	On the other hand, the tens of thousands of bits of
information needed to specify which pure state out of $e^S$ is
considerably more than the few hundred to few thousand bits of
information needed to specify which of the $N_{\mathrm vac} \sim
10^{100}-10^{1000}$ or so string vacua \cite{BP,Doug,AD,DD} we are
near.  Thus to say which pure state we are in, as well as what the
underlying vacuum is, might involve much fewer than a million bits of
information, assuming that the no-bang mixed state is correct and
that we focus on the states dominating the volume.  Publishing these
bits in a short paper would not be difficult if one had them, but it
is likely to be difficult to determine them.

	Here we have used the anthropic argument for large volume to
restrict the number of states enormously from the $e^{S_{\mathrm
no-bang}} \sim \exp{(4\times 10^{90})}$ number of states (for our
value of the cosmological constant; even much larger for vacua with
lower values) to $e^S \sim \exp{(10^{4.5})}$ volume-weighted
inflationary states if the no-bang mixed state is correct.  To
specify one out of all the first set of non-anthropic states would
require billions of times more information than even the entire
number of atoms in the observable part of our universe.  Since almost
all of these enormously many states do not have (rapid) inflation,
the no-bang quantum state is one example in which inflation is
extremely improbable when one considers all possibilities but is
highly probable (at least if eternal inflation can occur) when one
uses the anthropic selection effect of weighting by observers.

	A problem with the no-bang proposal is that for it to work,
one needs to invoke appropriate quantum fluctuations of the inflaton
to produce eternal inflation and yet the asymptotically de Sitter
regions produced by the much larger number of non-inflationary states
do not produce far more disordered observers or Boltzmann brains
\cite{Bolt,Rees,AS,PageBB,DP2006,BF,BBB,LindeBB,Vilfreak,Vanchurin,
BanksBB,Carlip,HS,TegBB,GM,McI} than the much smaller number of
inflationary states produce of ordered ordinary observers.  That is,
the no-bang proposal seems to suffer from the same qualitative
problem as the no-boundary proposal in very heavily weighting
non-inflationary solutions over inflationary solutions \cite{DP2006},
though for the no-bang proposal the quantitative problem is the
factor of `only' $e^S_{\mathrm no-bang} \sim \exp{(4\times 10^{90})}$
rather than the much bigger factor of $e^{S_{\mathrm dS}} \sim
\exp{(1.4\times 10^{122})}$ for the no-boundary proposal.

\section{Anisotropic necks}

	So far we have focused on seed states on necks that are
nearly homogeneous and isotropic.  However, Giddings and Marolf
\cite{GM} note that there may be far more nonperturbative states, of
the order of $e^{S_{\mathrm dS}} = \exp{(\pi b^2/G)} \sim
\exp{(1.4\times 10^{122})}$, where for the numerical values I am
using \cite{DP2006} the data of \cite{WMAP,Freedman}.  They cite the
conjecture \cite{Banks,Fischler} ``that the full space of
asymptotically (past and future) de Sitter states is of finite
dimension, with entropy given by the Bekenstein-Hawking value'' and
say ``we should include black holes (and possibly other
non-perturbative gravitational states) up to the maximum size $\sim
R$.''  Here I find that if one just requires nonsingular seed states
on a neck, the number of states can be infinite, but if one requires
that these states evolve to just a single asymptotically de Sitter
space (in both the future and in the past), the number of states may
be finite and of order $e^{S_{\mathrm no-bang}}$ rather than of the
order of the enormously larger $e^{S_{\mathrm dS}}$.

	A gravitational state that is a highly nonperturbative
deviation from de Sitter space is the Kottler \cite{Kottler} or
Schwarzschild-de Sitter metric
\begin{equation}
ds^2 = - \left(1-\frac{2M}{r}-\frac{r^2}{b^2}\right)dt^2
+ \left(1-\frac{2M}{r}-\frac{r^2}{b^2}\right)^{-1}dr^2
+ r^2(d\theta^2 +\sin^2\theta d\varphi^2),
\label{eq:10}
\end{equation}
which has both a cosmological constant $\Lambda = 3/b^2$ and a mass
parameter $M$ (but no other matter).  It has the topology of $R^1
\times R^1 \times S^2$ or $R^1 \times S^1 \times S^2$ if identified
to make it spatially compact, with homogeneous isotropic 2-spheres. 
A metric of this general topology and symmetry can be written as
\begin{equation}
ds^2 = - dt^2 + L^2(t,x)dx^2
+ r^2(t,x)(d\theta^2 +\sin^2\theta d\varphi^2),
\label{eq:11}
\end{equation}
where $x$ has either an infinite range (the $R^1 \times R^1 \times
S^2$ case) or is periodically identified (the $R^1 \times S^1 \times
S^2$).

	If such a metric has a neck, in the appropriate coordinate
system it would be a surface of constant $t$ (say $t=0$) at which
$\int r^2 L dx$ has a minimum.  One can always make a local rescaling
of $x$ so that $L(0,x) = 1$ at the neck.  Then the 3-metric of the
neck may be written as
\begin{equation}
ds^2_{\mathrm neck}
 = dx^2 + r^2(x)(d\theta^2 +\sin^2\theta d\varphi^2),
\label{eq:12}
\end{equation}
described by the single function $r(x)$ (and by the periodicity of
$x$ in the case in which the neck topology is $S^1 \times S^2$ rather
than $R^1 \times S^2$).

	In the generic Kottler or Schwarzschild-de Sitter metric, the
radius $r$ of the 2-sphere oscillates with the $R^1$ or $S^1$
coordinate $x$, taking values between the two positive roots of
$1-2M/r+r^2/b^2$.  The limiting case in which the two positive roots
coalesce (which does {\it not} make the event horizons at the two
roots coalesce, since the proper distance between them remains
finite, but only makes their areas the same), $M = 1/\sqrt{9\Lambda}
= b/\sqrt{27}$, leads to $r=1/\sqrt{\Lambda}=b/\sqrt{3}$ being
constant over the entire spacetime, and the resulting Nariai metric
\cite{Nariai} may be written in the form (\ref{eq:11}) as
\begin{equation}
ds^2 = - dt^2 + \cosh{(\sqrt{3}t/b)}dx^2
+ (b^2/3)(d\theta^2 +\sin^2\theta d\varphi^2).
\label{eq:13}
\end{equation}

	Unlike the generic Kottler or Schwarzschild-de Sitter, which
extends to $r=0$ and has a curvature singularity there, the Nariai
metric, which has $r$ constant everywhere, is completely
nonsingular.  However, it has no matter and hence no matter entropy. 
Therefore, let us add thermal radiation to the neck to see how large
we can make the matter entropy.  Again the condition that the neck is
a hypersurface of minimal 3-volume implies that the temperature $T$
of the thermal radiation must be less than the value given by Eq.
(\ref{eq:4}) and can attain that value in the case $\dot{\phi}=0$. 
For simplicity, let us assume that the neck is a moment of time
symmetry, a hypersurface of zero extrinsic curvature with
$\dot{\phi}=0$, and not just a hypersurface with the trace of the
extrinsic curvature is zero.

	The Einstein constraint equation for the neck with metric
(\ref{eq:12}) then gives
\begin{equation}
\frac{1-r'^2-2rr''}{r^2} = -G^0_0
= \Lambda + 4\pi G(m^2\phi^2 + 2aT^4),
\label{eq:14}
\end{equation}
where a prime denotes a derivative with respect to the $R^1$ or $S^1$
coordinate $x$.  Since the right hand side of this constraint
equation is positive, this equation can certainly be solved for
$r(x)$.

	In particular, when the right hand side is constant (as a
function of $x$), then there is the homogeneous solution $r =
[\Lambda + 4\pi G(m^2\phi^2 + 2aT^4)]^{-1/2}$.  If $x$ has length $X$,
then the total 3-volume of this homogeneous neck is $4\pi r^2 X$, and
the entropy of the thermal radiation is $S = (16\pi/3)aT^3r^2X$.  But
since the length X can be made arbitrarily large, the entropy of the
neck is unbounded above.

	That is, a positive cosmological constant does {\it not}
prevent the matter entropy, even on a regular minimal
hypersurface, from being infinitely large, because the minimal
hypersurface can be chosen to be infinitely large.

	This homogeneous example does not give an asymptotically de
Sitter space, since its time evolution causes $r$ to collapse to zero
at a big bang (in the past) and at a big crunch (in the future).  One
can see this for the metric (\ref{eq:11}), because of the Einstein
equation
\begin{equation}
-\frac{\ddot{r}}{r}-\frac{r''}{rL^2}
+\frac{\dot{r}\dot{L}}{rL}+\frac{r'L'}{rL^3}
= \frac{1}{2}\left(-G^0_0 + G^1_1\right)
= 4\pi G \dot{\phi}^2 + (16\pi G/3)aT^4,
\label{eq:15}
\end{equation}
assuming for simplicity on the far right hand side that $\phi$ is
independent of $x$, $\theta$, and $\varphi$ for each $t$.  The
homogeneous initial data gives $r'=0$ and $L'=0$, so
\begin{equation}
-\frac{L}{r}\left(\frac{\dot{r}}{L}\right)^\cdot
= 4\pi G \dot{\phi}^2 + (16\pi G/3)aT^4 > 0,
\label{eq:16}
\end{equation}
which causes $r$ to collapse to zero size in a finite time.

	On the other hand, we can choose inhomogeneous initial data
in which $r(x)$ oscillates (and the density of the thermal radiation
also oscillates appropriately to solve the constraint equation).  In
the regions where $r''$ is sufficiently negative (positive $r$
concave downward as a function of $x$), one can have $\ddot{r}>0$,
and such regions can expand indefinitely to become asymptotically de
Sitter regions in both the infinite past and in the infinite future. 
Again by having $X$, the range of $x$, large enough, one can get an
arbitrarily large amount of matter entropy on the neck.  Therefore,
even if one requires asymptotically de Sitter regions in both the
past and the future, a regular neck can have an infinite matter
entropy.

	One can also do this even if the topology of the neck is
$S^3$ rather than $R^1\times S^2$ or $S^1\times S^2$, simply by
choosing $r(x)$ on the neck to go to zero (with $r'=\pm 1$ to avoid
singularities) at the two ends of the range for $x$.  One can
still have an unbounded range of $x$ in between, and hence an
unbounded amount of matter entropy on the regular $S^3$ neck.

	Since the regions which expand into asymptotically de Sitter
space have $\ddot{r} > 0$, they must have $r''$ sufficiently negative
to solve the Einstein equation (\ref{eq:15}).  This gives a limit on
the range of $x$ which evolves to a single asymptotically de Sitter
region, and I would conjecture that the matter entropy on that part
of the neck is bounded by $S_{\mathrm no-bang}$ given by Eq.
(\ref{eq:7}).

	If one chooses an oscillating $r(x)$ with a greater range,
one will get a nested sequence of asymptotic de Sitter regions with
black hole singularities developing in between, rather than a single
asymptotic de Sitter region.  In particular, at points on the neck
where $r'=0$ and $r''>0$, one will have $\ddot{r}<0$, so at this
value of $x$ for both $t<0$ and for $t>0$ the 2-spheres of constant
$(t,x)$ will be closed trapped surfaces.  (Even if the resulting
black holes evaporate away by Hawking radiation, it does not seem
that this would eliminate the existence of more than one asymptotic
de Sitter region in the case in which one has the range of $x$ too
large at the neck.) 

	Therefore, if one limits the quantum states to give a regular
neck that evolves into only one asymptotically de Sitter region in
both the infinite past and the infinite future, as I have proposed
for the no-bang mixed quantum state of the universe, and if my
conjecture above is true, then the entropy of this state may indeed
be given approximately by $S_{\mathrm no-bang}$.

\section{Conclusions}

	In this paper I have proposed the no-bang mixed quantum state
for the entire universe or multiverse, which for each positive value
of the cosmological constant is an equal mixture of all quantum
states that may be generated by group averaging from pure seed states
that have a regular minimal hypersurface and which evolve to single
asymptotically de Sitter regions in the infinite past and future,
without any big bang or big crunch.  The total number of these pure
states that go into the no-bang mixed state is of the order of the
exponential of the three-quarter power of the Bekenstein-Hawking
entropy of de Sitter [in $D=4$ dimensions, otherwise the exponent is
not $3/4$ but $(D-1)/D$], at least if most of these pure states are
approximately homogeneous and isotropic on the largest scales.  This
number is enormously less than the number of quantum states normally
attributed to de Sitter.  The vast majority of even these restricted
no-bang states do not have rapid inflation (from an assumed inflaton
that can lead to exponential expansion much faster than the
asymptotic de Sitter regions).  However, if one includes an anthropic
selection effect by weighting by the 3-volume at some epoch after
possible eternal inflation, then a much smaller number of states may
dominate (much less than the apparent possible number of human
genomes) and make the resulting conditional probability of rapid
inflation very large.

	In this way an anthropic analysis of the no-bang quantum
state leads to inflation and the possibility of fitting our
observations of the universe (given a suitable inflaton, cosmological
constant, Standard Model of particle physics, etc., in at least some
part of the string or other landscape that is not atypical when
weighted by observations).  Of course, the no-bang state is just a
proposal that is not derived from deeper principles and admittedly at
present appears rather {\it ad hoc}, so it would be desirable to be
able to derive it (or an improvement) from simpler or more
fundamental principles.

\section*{Acknowledgments}

	I appreciated the hospitality of Edgar Gunzig and the
Cosmology and General Relativity symposium of the Peyresq Foyer
d'Humanisme in Peyresq, France, where Donald Marolf kindly and
patiently explained to me many aspects of quantum de Sitter space. 
Andrei Linde and Tom Banks have provided helpful comments on an
earlier version of the manuscript.  This research was supported in
part by the Natural Sciences and Engineering Research Council of
Canada.

\end{document}